\documentclass{PoS}

\usepackage{amsmath}
\usepackage{xspace}
\usepackage{setspace}
\usepackage{mathrsfs}

\newcommand{\jpsi}{J$/\psi$\xspace}
\newcommand{\psip}{$\psi(2\text{S})$\xspace}
\newcommand{\pt}{p_{\text{T}}}

\title{
	Charmonium production in pp collisions with ALICE at the LHC
}
\ShortTitle{
	Charmonium production at the LHC
}
\author{
	\speaker{Lucas Altenkamper}\thanks{for the ALICE collaboration}\\
        University of Bergen, Norway\\
        E-mail: \email{lucas.altenkamper@cern.ch}
}
\FullConference{
	The 39th International Conference on High Energy Physics (ICHEP2018)\\
	4-11 July, 2018\\
	Seoul, Korea
}

\abstract{
	Charmonia, bound states of charm and anti-charm quarks, represent an interesting probe for the study of Quantum Chromodynamics (QCD) since their production involves both hard and soft energy scales.
	Several effective models are available to describe the production of charmonia, but so far none have been able to describe all experimental observables simultaneously.
	ALICE has studied the production of charmonia in different collision systems at all available LHC energies at both mid- and forward-rapidity down to zero transverse momentum.
	In this contribution, different measurements performed in proton--proton (pp) collisions are presented, namely $\pt$-differential cross sections of inclusive \jpsi and \psip, inclusive \jpsi polarization measurements and the latest results on correlations between inclusive \jpsi and unidentified charged hadrons. 
	The results are also compared to model predictions.
}

\begin{document}

\section{Introduction}
The production mechanisms of charmonia  operate at the boundary of the perturbative and non-perturbative regimes of QCD.
The initial charm pair is produced in hard scattering processes where a perturbative description applies.
The evolution into a bound state is a non-perturbative process, which can only be described using effective models such as the Color Singlet Model (CSM) \cite{csm}, the Color Evaporation Model (CEM) \cite{cem} or Non-Relativistic QCD (NRQCD) \cite{nrqcd}.
Although charmonia have been studied extensively, their production mechanism is still not fully understood.
Measurements in pp collisions at the highest LHC energies can provide further constraints for prompt charmonium production models, contributing to a deeper understanding of the corresponding production processes.
\par
In collider experiments, charmonia are typically separated into two groups according to their origin.
Prompt charmonia are produced directly in the initial collision or, e.g. in case of the \jpsi, via feed down decays of heavier states such as the \psip or $\chi_c$.
Non-prompt charmonia originate in the weak decay of \textit{B} hadrons and thus exhibit a production vertex that is displaced from the collision vertex.
Inclusive charmonia refer to all charmonia without distinction of their production process.
\par
ALICE has measured the production of inclusive charmonia in pp collisions at all available LHC energies.
We present a summary of the most recent results on charmonium production and compare them to model predictions.

\section{Charmonium measurements with ALICE}
ALICE is the dedicated heavy ion experiment at the LHC.
A detailed description of the ALICE detector can be found in reference \cite{alice}.
Charmonia are reconstructed in two different rapidity regions: the decay into di-electrons is reconstructed at mid-rapidity ($|y|<0.9$) in the central barrel and the decay into di-muons is reconstructed at forward-rapidity ($2.5<|y|<4$) in the muon spectrometer.
The central barrel detectors relevant for charmonium measurements are the Inner Tracking System (ITS) and the Time Projection Chamber (TPC), with the latter being used for tracking and particle identification.
The muon spectrometer is screened by an absorber of ten interaction lengths and tracks muons in five multiwire proportional chambers.  
Two layers of resistive plate chambers in the muon spectrometer provide triggering for single- or di-muon events.
Additionally, the two V0 scintillator arrays, situated at $-3.7<\eta<-1.7$ and $2.8<\eta<5.1$, are used for triggering.
\par
Charmonium candidates are reconstructed from their di-lepton decay channels.
Oppositely charged tracks reconstructed in the central barrel (TPC+ITS) or the muon spectrometer are combined into pairs and the charmonium signal is extracted from the invariant mass distribution of the pairs.
In the central barrel, electron tracks are identified using their specific energy loss in the TPC while the tracks reconstructed in the muon spectrometers are identified as muons.

\section{Results}

\begin{figure}[htb]
	\center
	\includegraphics[width=.45\textwidth]{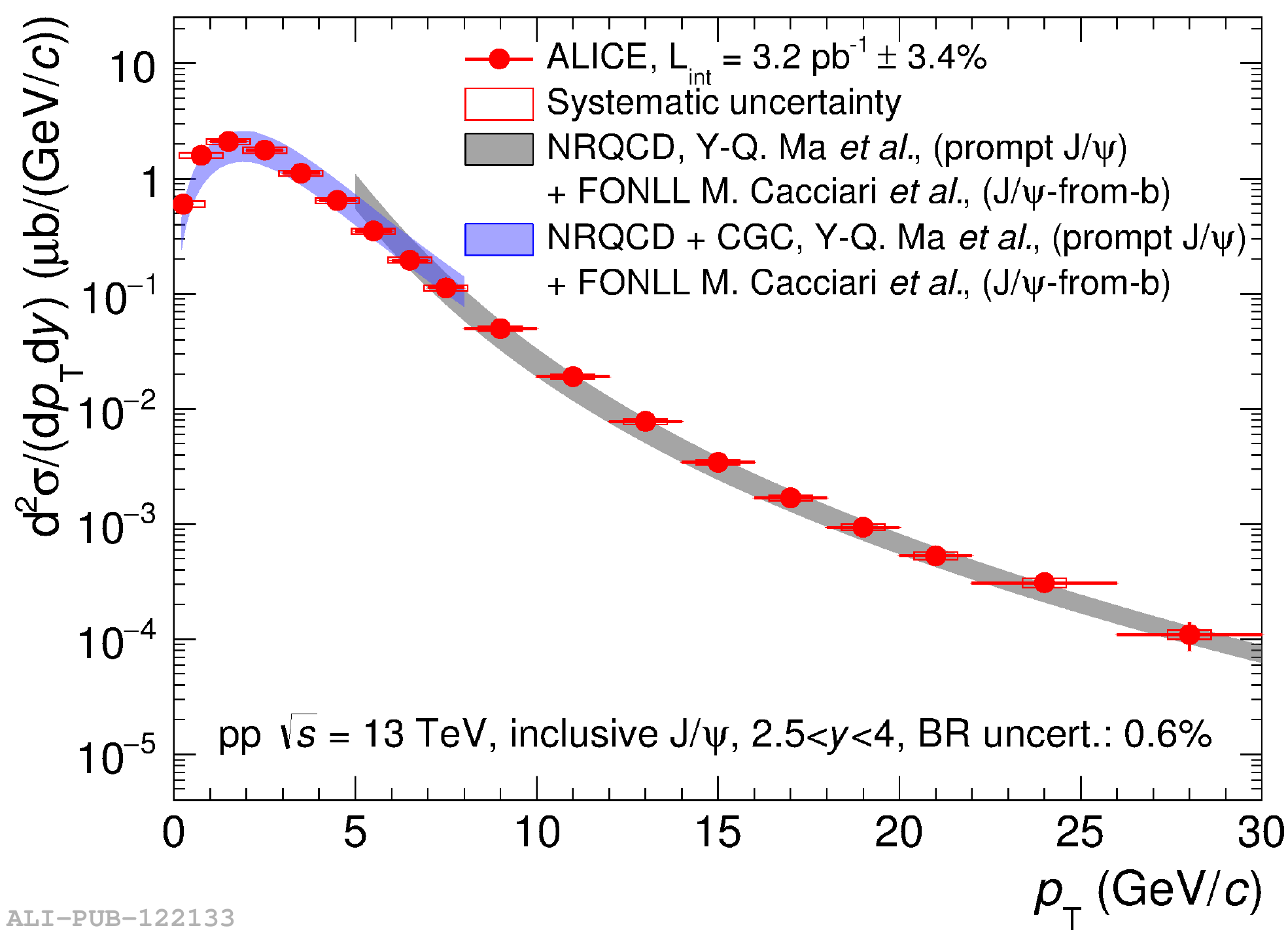}
	\hspace{10pt}
	\includegraphics[width=.45\textwidth]{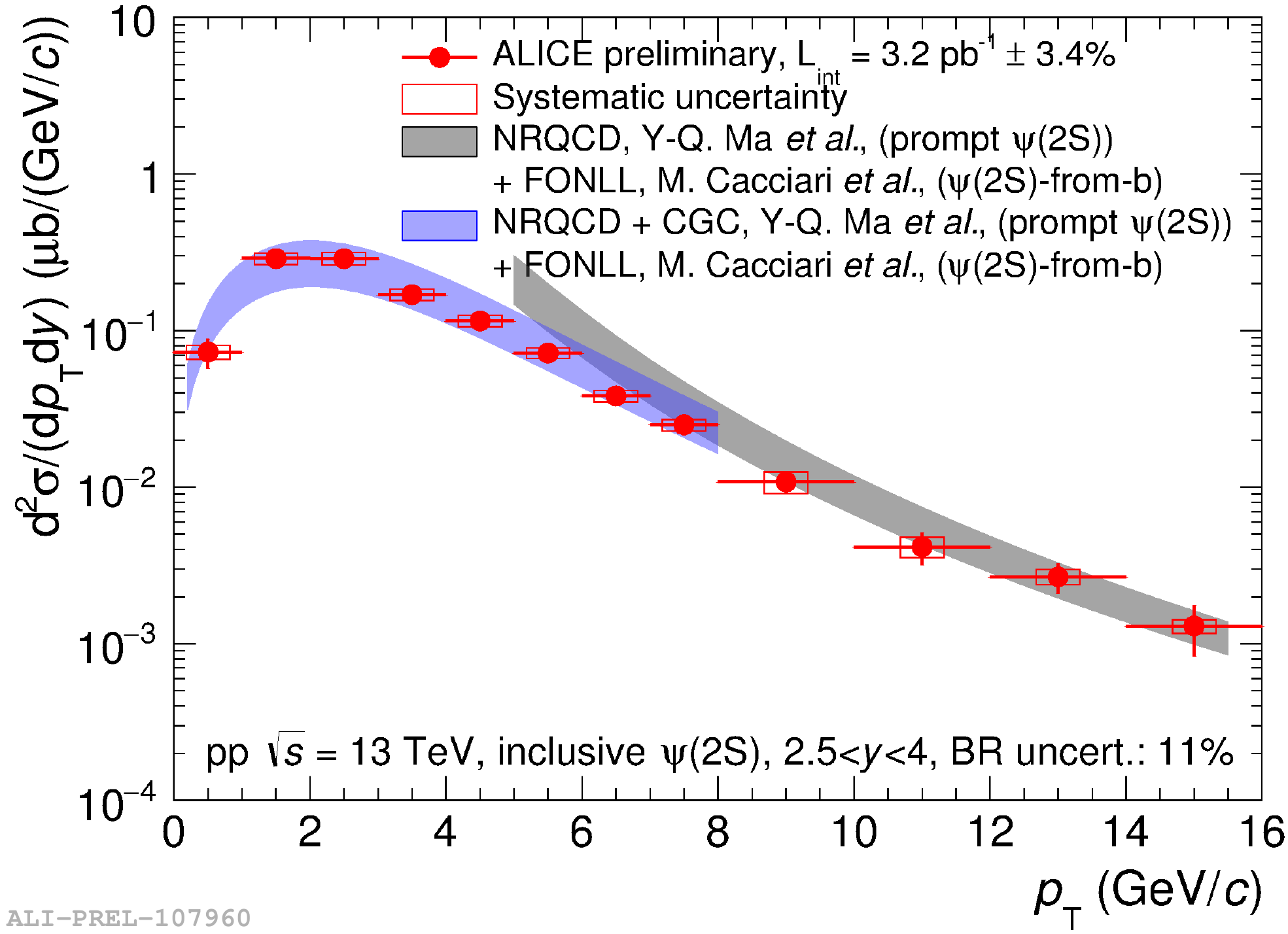}
	\caption{Inclusive \jpsi (left) and \psip (right) cross sections at forward-rapidity in pp collisions at $\sqrt{s} = 13~\text{TeV}$, compared to the sum of model predictions for the prompt and non-prompt cross sections.}
	\label{fig:crosssection}
\end{figure}

\paragraph{\normalfont\itshape Inclusive \jpsi and \psip cross section} 
ALICE has measured the inclusive \jpsi and \psip cross sections in pp collisions at forward-rapidity at different center-of-mass energies \cite{dimuon}. 
Results at $\sqrt{s} = 13~\text{TeV}$ are presented in figure \ref{fig:crosssection}.
The cross section measurements are compared to an ad-hoc model of the inclusive cross section, which represents the sum of the prompt and non-prompt contributions.
Hereby, the prompt cross section is described by a NRQCD \cite{promptcross} (NRQCD coupled with Color Glass Condensate (CGC) \cite{promptcgccross}) model at high (low) $\pt$.
The non-prompt cross section is described by FONLL \cite{nonpromptcross}. 
A fair agreement between the model predictions and the data is observed for both \jpsi and \psip.

\begin{figure}[htb]
	\center
	\includegraphics[width=.45\textwidth]{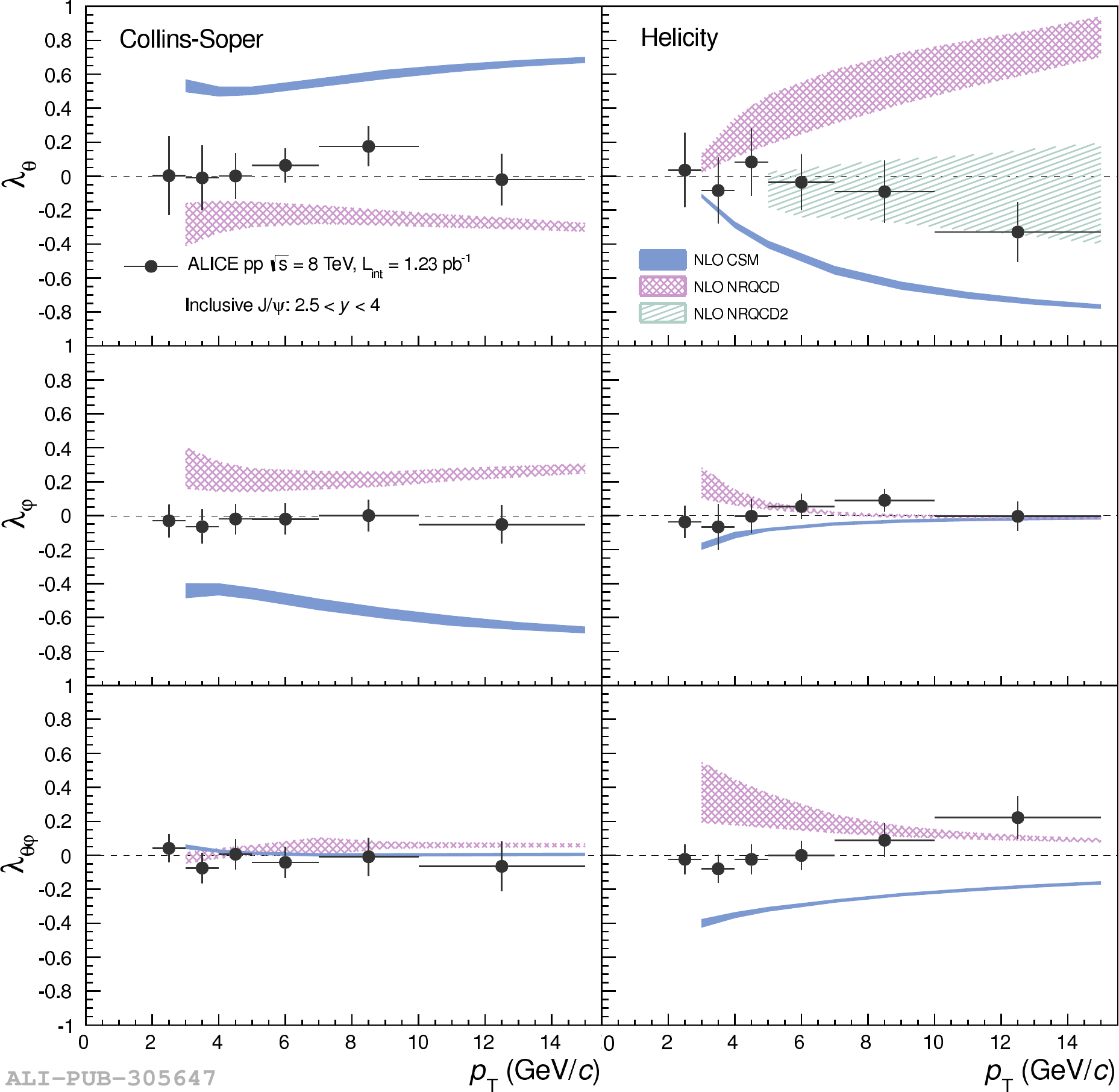}
	\hspace{10pt}
	\includegraphics[width=.45\textwidth]{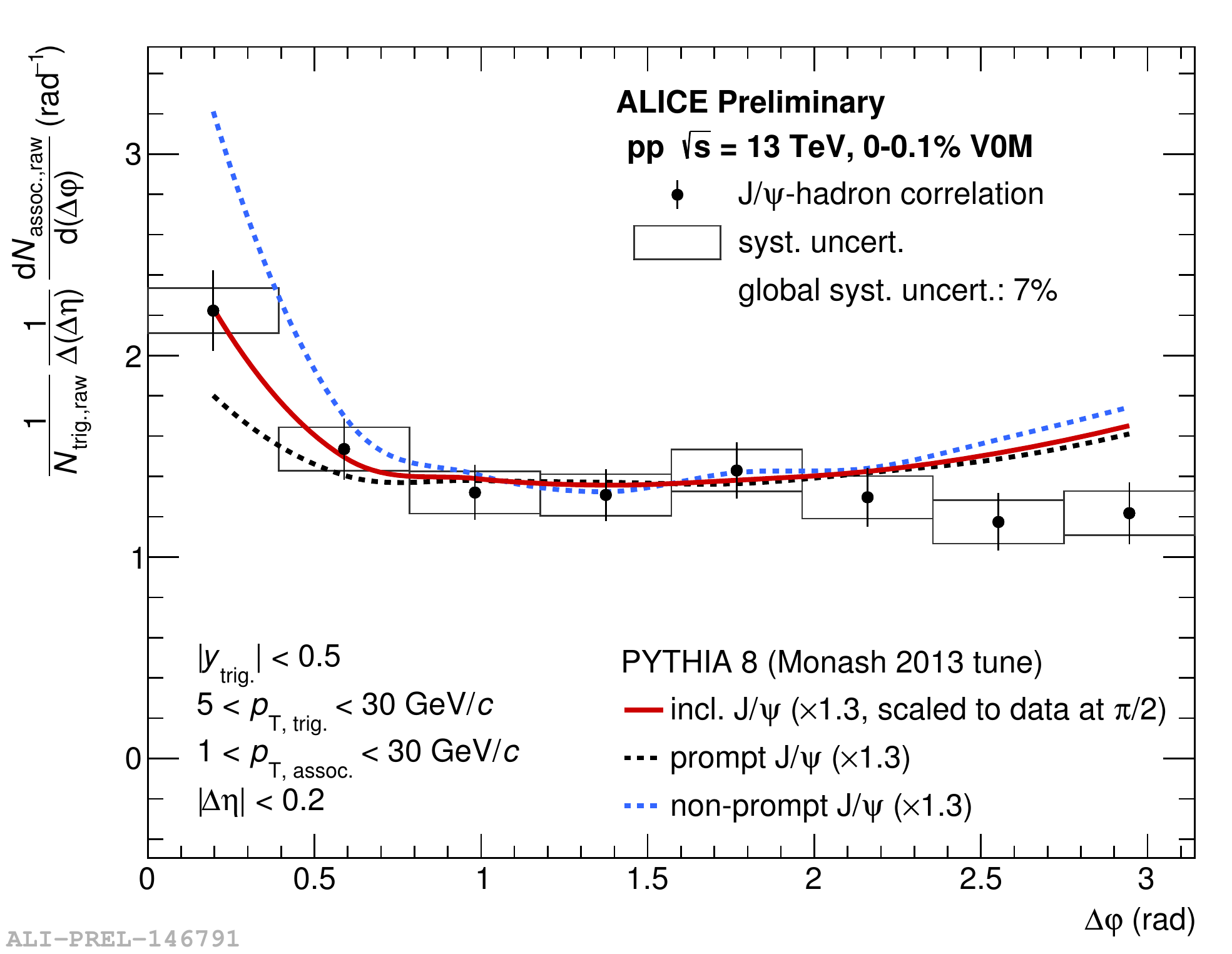}
	\caption{\textbf{(left)} Inclusive \jpsi polarization measured at forward-rapidity in pp collisions at $\sqrt{s} = 8~\text{TeV}$, compared to model predictions from CSM and NRQCD for prompt \jpsi. \textbf{(right)} Inclusive \jpsi-hadron correlations measured at mid-rapidity in pp collisions at $\sqrt{s} = 13~\text{TeV}$, compared to PYTHIA 8.}
	\label{fig:polAndCorr}
\end{figure}

\paragraph{\normalfont\itshape Inclusive \jpsi polarization}
The polarization of inclusive \jpsi has been measured by ALICE at forward-rapidity in pp collisions at $\sqrt{s} = 7~\text{TeV}$ \cite{pol2} and $\sqrt{s} = 8~\text{TeV}$ \cite{pol} in the Collins-Soper and helicity reference frames.
The polarization parameters measured in pp collisions at $\sqrt{s} = 8~\text{TeV}$, determined from the angular distribution of the muon pair from the di-muon decay channel, are shown in the left panel of figure \ref{fig:polAndCorr} for both frames.
The measurement is in agreement with no polarization for inclusive \jpsi, and lies between the CSM \cite{csmpol} and NRQCD \cite{csmpol,nrqcdpol} model predictions.
However, it should be noted that the polarization of inclusive \jpsi is compared to model predictions for prompt \jpsi.

\paragraph{\normalfont\itshape Inclusive \jpsi-hadron correlations}
A preliminary result on the correlation between inclusive \jpsi and unidentified charged hadrons measured at mid-rapidity in pp collisions at $\sqrt{s} = 13~\text{TeV}$ is shown in the right panel of figure \ref{fig:polAndCorr}.
The correlation is expressed in terms of a raw per-trigger yield as a function of the angular distance $\Delta\varphi$, which is not yet corrected for the detector efficiency.
A near-side peak can be observed while the away-side peak is diluted, partly due to a sharp cut on the distance in pseudorapidity ($|\Delta\eta|<0.2$).
The data is compared to PYTHIA 8 \cite{pyth}, where contributions from prompt and non-prompt \jpsi are shown separately in addition to the distribution for inclusive \jpsi.
The comparison suggests that the near-side correlation is mainly driven by the non-prompt contribution, i.e. either by additional decay products in the vicinity of the \jpsi or by fragments from the hadronization of the initial b quarks.

\section{Summary}
We have shown the most recent results on the production cross sections of inclusive \jpsi and \psip, which are in fair agreement with the model predictions. 
A measurement of the inclusive \jpsi polarization has been presented, which provides new constraints on the model predictions for prompt \jpsi.
Additionally, a preliminary result on the correlation between inclusive \jpsi and unidentified charged hadrons was shown and compared to PYTHIA 8.
The available models show difficulties in describing all experimental observables simultaneously.
An expansion of the correlation measurement to prompt \jpsi in the future might help to constrain model predictions further.

{
\setstretch{0.9}

}


\begin{thebibliography}{99}
	\footnotesize
	\setlength{\itemsep}{0pt plus 0.3ex}
	\bibitem{csm} 
		H. Fritzsch, 
		\emph{Producing heavy quark flavors in hadronic collisions - A test of quantum chromodynamics},
		\emph{Phys. Lett. B} 67 (1977) 217
	\bibitem{cem} 
		M. B. Einhorn, S. D. Ellis, 
		\emph{Hadronic production of the new resonances: Probing gluon distributions},
		\emph{Phys. Rev. D} 12 (1975) 2007
	\bibitem{nrqcd} 
		G. T. Bodwin \textit{et al.}, 
		\emph{Rigorous QCD analysis of inclusive annihilation and production of heavy quarkonium},
		\emph{Phys. Rev. D} 51 (1995) 1125, \textit{errata} 55 (1997) 5853 
	\bibitem{pol2}
		B. Abelev \textit{et al.} (ALICE Collaboration),
		\emph{\jpsi polarization in pp collisions at $\sqrt{s} = 7~\text{TeV}$},
		\emph{Phys. Rev. Lett. 108 (2012) 082001},
		{\tt arXiv:1111.1630 [hep-ex]}
	\bibitem{pol} 
		S. Acharya \textit{et al.} (ALICE Collaboration), 
		\emph{Measurement of the inclusive \jpsi polarization at forward rapidity in pp collisions at $\sqrt{s} = 8~\text{TeV}$},
		\emph{Eur. Phys. J. C} 78 (2018) 562,
		{\tt arXiv:1805.04374 [hep-ex]}
	\bibitem{alice} 
		K. Aamodt \emph{et al.} (ALICE Collaboration), 
		\emph{The ALICE experiment at the CERN LHC},
		\emph{JINST} 3 (2008) S08002 
	\bibitem{dimuon} 
		S. Acharya \emph{et al.} (ALICE Collaboration), 
		\emph{Energy dependence of forward-rapidity \jpsi and \psip production in pp collisions at the LHC},
		\emph{Eur. Phys. J. C} 77 (2017) 392,
		{\tt arXiv:1702.00557 [hep-ex]}
	\bibitem{promptcross} 
		Y-Q. Ma \emph{et al.}, 
		\emph{\jpsi$(\psi')$ Production at the Tevatron and LHC at $\mathscr{O}(\alpha_s^4v^4)$ in Nonrelativistic QCD},
		\emph{Phys. Rev. Lett.} 106 (2011) 042002,
		{\tt arXiv:1009.3655 [hep-ph]}
	\bibitem{promptcgccross} 
		Y-Q. Ma \emph{et al.}, 
		\emph{Comprehensive description of \jpsi production in proton-proton collisions at collider energies},
		\emph{Phys. Rev. Lett.} 113 (2014) 192301,
		{\tt arXiv:1408.4075 [hep-ph]}
	\bibitem{nonpromptcross} 
		M. Cacciari \emph{et al.}, 
		\emph{Theoretical predictions for charm and bottom production at the LHC}
		\emph{J. High Energ. Phys.} 1210 (2012) 137
	\bibitem{csmpol} 
		M. Butenschoen, B. A. Kniehl, 
		\emph{\jpsi Polarization at the Tevatron and the LHC: Nonrelativistic-QCD Factorization at the Crossroads},
		\emph{Phys. Rev. Lett.} 108 (2012) 172002
	\bibitem{nrqcdpol} 
		K-T. Chao \emph{et al.}, 
		\emph{\jpsi Polarization at Hadron Colliders in Nonrelativistic QCD},
		\emph{Phys. Rev. Lett.} 108 (2012) 242004
		{\tt arXiv:1201.2675 [hep-ph]}
	\bibitem{pyth} 
		P. Skands, S. Carrazza, J. Rojo, 
		\emph{Tuning PYTHIA 8.1: the Monash 2013 Tune},
		\emph{Eur. Phys. J. C} 74 (2014) no. 8, 3024,
		{\tt arXiv:1404.5630 [hep-ph]}
\end{thebibliography}
\end{document}